# Some Aspects of Classical and Quantum Phases

Yakir Aharonov [1], Tirzah Kaufherr, Shmuel Nussinov[1,2]

*Tel Aviv University, Sackler Faculty of Sciences, Ramat Aviv, Tel Aviv 69978, Israel*
*(1)Chapman University, Schmid College of Science, Dept. of Physics, Computational Sciences and Engineering, One University Drive, Orange CA 92866, USA*
*(2)E-mail: nussinov@post.tau.ac.il*
*Foreword: This is based on a lecture given by S.N. on the occasion of the 75th birthday of F.T. Avignone and E. Fiorini in Columbia, South Carolina.*

## Abstract

We study classical and quantum phases in the adiabatic Born-Oppenheimer context. These include a classical astronomical case, the general dual description of the phases, a new "Paradox" connected to scattering Berry phase and its resolution and various elaboration of topological/geometrical/non-abelian phases.

# 1 Introduction

Adiabatic Born-Oppenheimer settings are common in classical and quantum physics. Thus envision a point-like massive object (or strongly charged particle)—the "Strong Probe"—passing by a binary star (or an atom)—the "System". If the probe's motion is slow compared with characteristic periods of the system (or $h/(E_i - E_0)$ for the atom in a non-degenerate ground state), we can use an adiabatic approximation and solve the dynamics of the system—or the Schrödinger's Equation—for each "instantaneous" location $R(t)$ of our probe. The resulting energy shift $E_0(R) - E_0(R = \infty)$ defines the effective "B.-O. potential" that our external probe sees. Interfering parts of a quantum mechanical wave function following different paths allows sensing potential differences even in cases where the test particle is always in force-free regions.

In the following we focus on the general principle that phases be understood from both the point of view of the external probe particle and that of the dynamical system with which it interacts. This principle affords a unified point of view of Q.M. phases yielding, in particular, a new "Paradox" for scattering type adiabatic set-ups and a very amusing resolution thereof.

## 2 A classical (astronomical) Berry phase

The adiabatic approximation was used by Gauss for celestial motions. Also here we have on top of the dynamical effect of changed period an extra Berry phase (or time lapse). The latter resembles a spin-half dynamical system which attempts tracking the location of the external probe, and the accumulated effect of it failing to do so instantaneously is the analog of the geometric phase. (See fig. 1 for the general adiabatic setting.)

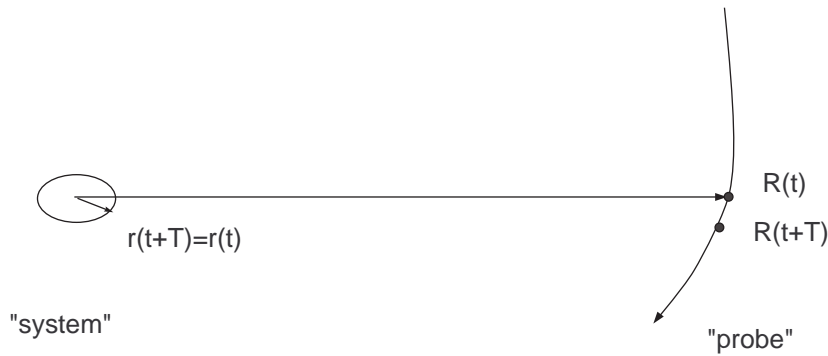

Fig. 1. Adiabatic setting

Consider a test mass (Earth) moving around a very heavy fixed center (Sun) with another heavy planet (Jupiter) present. We will assume that Jupiter's orbit is coplanar with Earth's orbit. The question of interest is how does the presence of Jupiter effect the period (year) of the earth? Since $M_{\text{Jupiter}} = 10^{-3} \, M_{\text{Sun}}$, $R_{\text{Jupiter}} = 5.2 \, R_{\text{Earth}}$ and $T_{\text{Jupiter}} = 11.9 \, T_{\text{Earth}} = 11.9$ years, both a perturbative (in force ratio $F_{\text{Jupiter}}/F_{\text{Sun}} \sim 5.10^{-5}$) *and* adiabatic (in the sense that $T_{\text{Jupiter}} > T_{\text{Earth}}$) approaches are feasible. The latter suggests solving the problem for an instantaneous frozen position of Jupiter. In particular, we need the new period $T'_{\text{Earth}}$ in the presence of the Sun and Jupiter. Remarkably, that special planar two-center problem is exactly solvable. (One can also use perturbative approximation in this case.) The center problem with all three bodies in a fixed plane has been solved analytically by Liouville. See [2].

There is, however, a finite extra correction; namely, the Berry-like delay—the very analog of the geometric Berry phase arising from the fact that the dynamical system, the orbiting earth, fails to instantaneously adjust itself (the orbit's period in the case of interest here) to the new position of the probe (Jupiter)—the sum of these over one Jupiter period, 'the astronomical Berry phase' can be computed.[3] Unlike the huge ordinary Gauss/Born-Oppenheimer effect, it is barely observable.

# 3 The two complementary ways of viewing quantum phases

When the external probe describes a closed path and the dynamical system returns to its initial position, its wave function (and that of the system as a whole) can only pick up a phase. It is very instructive to understand how this phase arises from two complementary points of view: that of the motion and interactions of the external probe alone or, alternatively, by considering the "Internal" dynamics of the system.

This applies to all cases. For the electric (or magnetic) potential effects the A.B. phase emerges via relative linear momentum (or angular momentum) imparted by the electron to the plates (solenoid). This is particularly transparent in the electric case. (see fig. 2.)

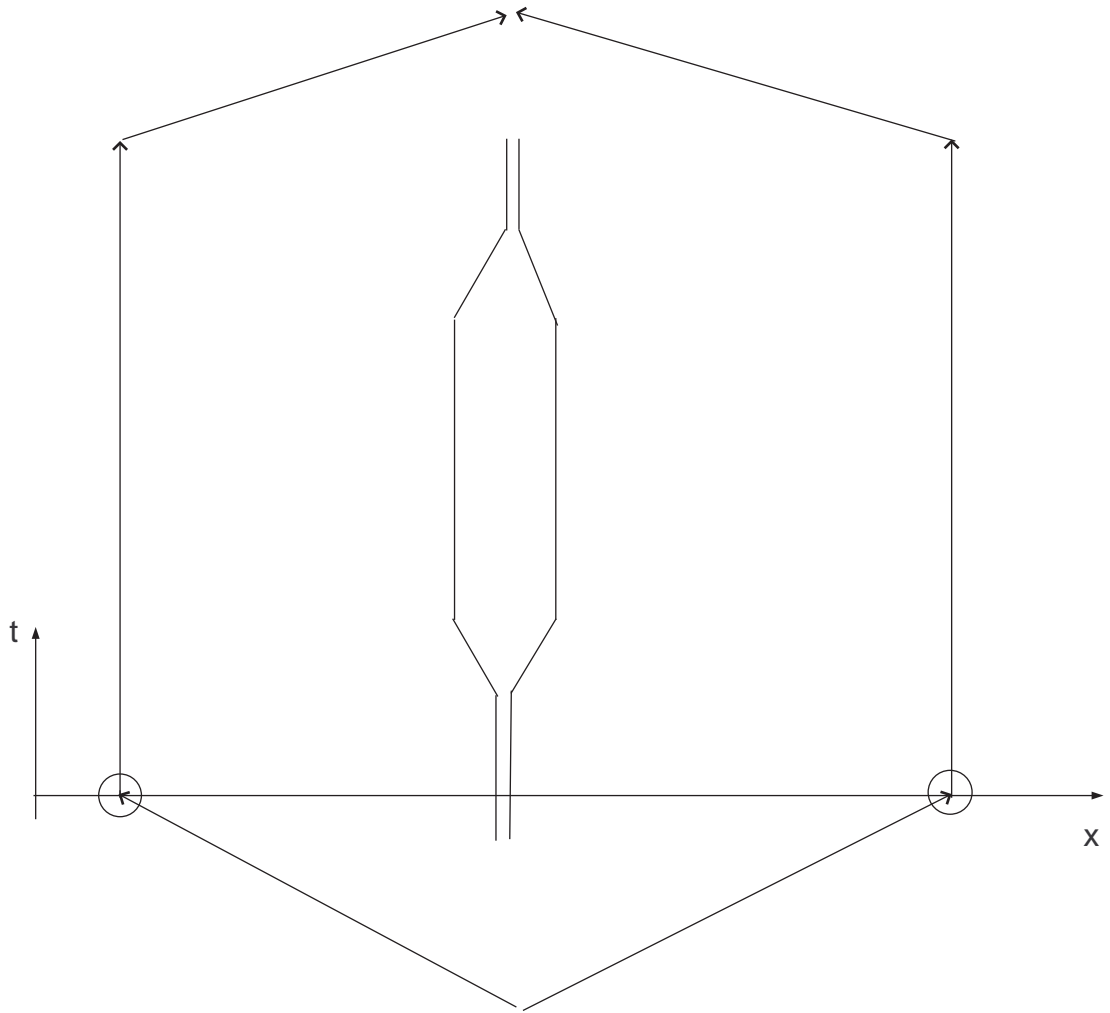

Fig. 2. The Aharonov Bohm Electric Effect

An electron, playing here the role of an external *weak* probe, stays for time $t$ in the force-free region outside two parallel condenser plates. During this time the plates are at a fixed relative distance $x$ with a uniform field $E$ existing between them. The standard (electric) potential modifies the wave function of an electron passing to the left-L of the plates by the phase $eV_L \cdot t$ = eExt (or $eV_R \cdot t$ = -eExt for electrons passing on the right. The relative A.B. phase 2eExt then manifests in the interference of these two paths.

The forces $F_{L,R}$ = +/− eE between the electron and the L/R plate, cancel out for the electrons and for the motion of the center of mass of the whole capacitor—i.e., the two plates jointly. However, insofar as the *relative* motion of the two plates is concerned (i.e., the internal dynamics of the "system" in the present case), the effects add up: $F_L$ imparts a momentum $P_x(L) = Ft$ =eEt to the left plate and $F_R$ imparts the opposite momentum $P_x(R)$ = -eEt to the right plate at a relative distance $x$ away. The phase picked up by the internal (relative $x$) wave function of the system is thus 2eEt·$x$—the precise A.B. phase.

Localizing the electron to the left *or* to the right of the capacitor—namely, within $\Delta(X) < x$, induces by the uncertainty principle an uncertainty $\Delta(P_x) > 1/x$ in its momentum. The A.B. phase $\sim$ eExt can be taken as $< \pi$. The uncertainty in the imparted momentum does then exceed the momentum itself. Consequently the A.B. phase becomes uncertain and, as expected, knowing which path (L or R) the electron took destroys the interference.

For complex "Systems" the task of identifying the internal dynamical change accounting for the phase—otherwise simply explained by an appropriate (albeit somewhat mysterious at times) external potential, becomes increasingly difficult. Already for the magnetic A.B. effect, one needs to use a simplified "solenoid"—say, counter-rotating almost overlapping concentric cylinders with opposite charges—in order to have the analog of the above simple reasoning as in the electric A.B. case. Still, the general principle that such alternative, complementary, interpretations of the phase are possible is quite powerful.

In Sects. 4 and 5 we briefly discuss two examples illustrating this. Implementing the above general principle appear at times to lead to a "Paradox". Such a case is presented along with its amusing resolution. The second example involves a discussion of well-known Berry phases.

# 4 A paradox related to scattering Berry phase and its resolution

The "system" with which our external weak and or strong probe interacts need not be static or bound as often implicitly assumed, but can involve a scattering state.

Thus let our system consist of a quantum particle (electron, photon, neutron, etc.) incident from the right ($x = +\infty$) along the $x$ axis, moving in a semi-infinite channel and reflecting from a mirror at the origin ($x = 0$). The small channel width $w$, prevents excitation of transverse ($y$) modes rendering the system one dimensional.

The particle's wave function is then

$$\psi(x) \sim \exp(-ipx) - \exp(ipx) \tag{1}$$

so that it vanishes at $x = 0$, the location of our mirror. In reality we can have an incident wave packet, which encounters the mirror at time $t \sim 0$ and becomes the reflected wave packet. Our external, strong probe is located outside the channel at $(X, Y)$, $X > 0$ and $Y > w$. Its short-range repulsive interaction with the internal particle $U(X, Y; x) = U(|r - R|)$ is strong at short distances $|r - R| = [(x - X)^2 + Y^2]^{1/2} <$ few $w$, and decreases to almost zero for $|r - R| > Y_0$. Hence when at small $Y$ the probe reflects the internal particle at $x = X$ before it reaches the origin introducing an extra phase $\sim 2pX$. On the other hand, when the probe is far from the channel at $Y \sim Y_0$ it will not effect the incident internal particle which will reflect from the mirror with no extra phase. A heavy probe which hardly moves during the collision will pick (after a time $T \sim Y_0/v$ with $v = p/m$ the velocity of the internal particle required for the latter to reflect and stop interacting) a phase $\Phi(X, Y)$.

The overall change of the phase as $Y$ changes between $w$ and $Y_0$ $\Delta \Phi \sim 2pX$ implies a phase gradient or momentum in the $y$ direction of $2pX/Y_0$ and the average force obtained by dividing by the effective duration of the collision, $T$, is $F \sim 2pXv/Y_0^2 \sim 2p^2X/mY_0^2$. However, it is clearly impossible for the force to indefinitely increase with the distance $X$ from the distant mirror at the origin.

The paradox arises due to a serious omission. Even when the external probe is very near to the $x$ axis the internal particle does not see an infinite barrier. Consequently there is some small probability, $\sim \epsilon$, that the particle will not reflect from the probe at $x = X$, imparting to it a momentum $2p$, but rather tunnel through the barrier. It will then be reflected from the mirror at $x = 0$ and subsequently by symmetry, will reflect back with a high probability, $1 - \epsilon$ from the left end of the barrier imparting to it a momen-tum $-2p$ (with probability $\epsilon \cdot (1-\epsilon)$). This backward and forward reflections keep going on. The n$^{th}$ reflection with probability $f_n = \epsilon \cdot (1-\epsilon)^{(n+1)}$ and will then impart a momentum $-2pf_n$. The geometric series sum of all the many small negative momentum kicks exactly cancels the original positive $2p(1-\epsilon)$. In physical terms the particle will be trapped, albeit with a small probability $\epsilon$ on the left-hand side of the barrier. The trapping, however, lasts for a long time, $\sim 1/\epsilon$, thereby equalizing the pressures exerted on the right- and left-hand side of the barrier (namely, the external probe) at $x = X$. Thus our probe suffers no net momentum kick and the internal particle reflects just from the end mirror as if nothing much happened in between!

The latter suffer no net momentum kick and the internal particle reflect just from the end mirror as if nothing much happened in between!

Is it possible to have an actual realization of this using laser beam of light and a tiny mirror? In such a case one could envision reflection by the probe of one transverse polarization and ensuing extreme sensitivity to tiny rotations of the plane of polarization of the trapped photons during their many traversals. We will not speculate any further on this here. Experiments of this type are, in fact, being done at MIT by Nervis Paluvalaha and her group in connection with the ultimate quantum noise limitations on the measurements of gravitational waves utilizing multiple reflections of laser light from mirrors.

## 5 Berry-A.B. geometric/topological phases

Returning to the initial external slow probe polarizing an atom, consider the case of just a two-level system-"Atom" arising when we have not just one iso-lated ground state but two, nearby states well separated from all other levels.

The atom can then be replaced by a spin-half system at the origin and its interaction with the probe is $H = \vec{A} \cdot \hat{\sigma}$ with the vector consisting of the three Pauli matrices, appropriate for spin-1/2. For time reversal invariant interaction and no external B fields, only the real $\sigma(1)$ and $\sigma(3)$ appear.

The two-level spin system becomes degenerate when $|A| = 0$. For the time reversal invariant case this reduces to

$$A_1(\vec{R}) = A_3(\vec{R}) = 0. \tag{2}$$

Two equations define a one-dimensional closed (or infinite) curve C*. A basic result—reminiscent of the magnetic A.B. effect with a singular B field vortex—is the following: When the probe adiabatically describes a closed circuit C and the system returns to its initial non-degenerate ground state it picks a topological Berry phase, $\Phi_{Berry} = \pi$, if and only if the curves C and C* interlock.

Generally an $A_2 \, \sigma(2)$ term is present. Requiring that $A_2$ vanish as well reduces the degeneracy manifold in $R$ space to a single point $R^*$. In the famous example described next, this point $R^*=0$ is at the location of the system-"Atom" itself. A "Geometric" Berry phase manifests for probe system interaction: $A \, \hat{\sigma} \cdot \hat{n}(t)$ with $A$ a constant and $\hat{n}(t)$ the unit vector pointing towards $\hat{R}(t)$—the instantaneous probe's location. At the origin $\vec{R} = 0$ and hence the direction $\hat{n}$ and the Hamiltonian $H = A\hat{s} \cdot \hat{n}$ are ill defined. We define $H$—via a procedure similar to Schwinger's point splitting—by averaging the interaction over a spherical neighborhood around each $\vec{R}$. This does not change $H$ for any $|R| > 0$ but makes $H = 0$ at $R = 0$. Hence $\vec{R}^*=0$ as claimed above.

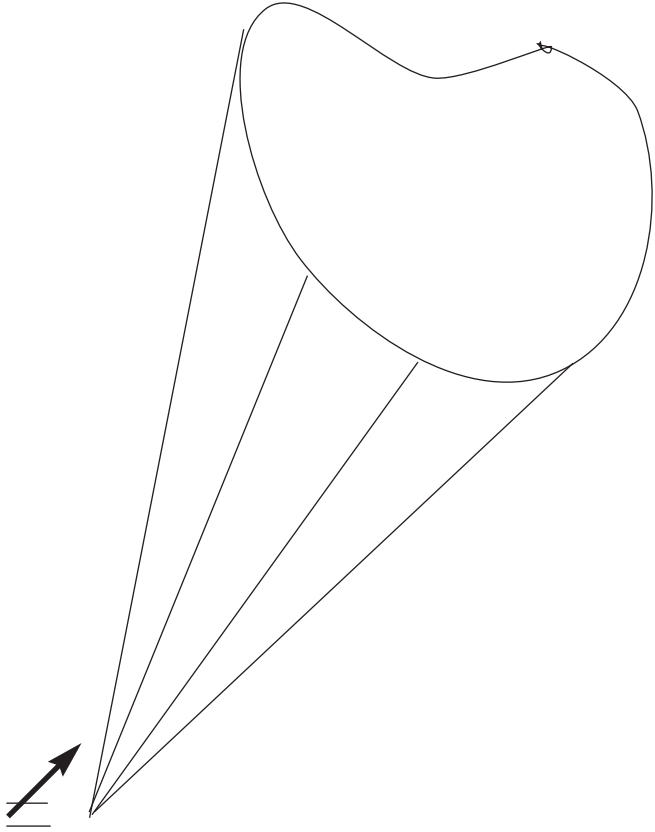

Fig. 3. Geometric Berry phase

As $\vec{R}(t)$ moves around the closed pathe of fig.3 the energy of the system, its interaction energy, $\sim |A|$, does *not* change and there are no B.O. forces. Yet the probe picks up after describing closed curve C a "Geometric phase" equal to half the solid angle substended by C at the origin. This and the independence of the phase on $|A|$, the interaction strength seems surprising!

We next consider this phase from both points of view: That of the probe and of the two-level system.

The first is well known and elaborated in many reviews: There *is* a "Lorentz"-like force acting on the probe which, however, is at all times perpendicular to its velocity so that no work is done and no energy change incurred. This force requires introducing a fictitious monopole of strength $g$ at the origin, endowing our probe with a fictitious charge $e$ and imposing Dirac quantization, $e \cdot g \sim h/2$. The standard A.B. phase, $\Phi_{A.B.} = \int_C e\vec{A} \cdot \vec{d\ell}$ with $\vec{A}$ the vector potential, is then the flux of the monopole's $\vec{B} = g\hat{r}/\vec{r}^2$ through surfaces bounded by $C$, namely, the solid angle above.

Note that this geometric Berry phase due to the fictitious monopole nicely ties in with the previous topological "A.B.-like" Berry phase. Thus consider imposing on the Hamiltonian $A \cdot \hat{n}$ the condition $\vec{A} \cdot \hat{n}_0 = 0$ for a specified direction $\hat{n}_0$. In this case H vanishes only on the infinite line directed along $\pm\hat{n}_0$. This line then becomes the degeneracy manifold—the curve C* mentioned above. A Probe encircling this line collects a topological Berry phase—which in terms of the formal e.m. analogy becomes an "A.B." phase due to a "fluxon" aligned along $\pm\hat{n}_0$. It is as if some symmetry breaking phase transition, like in superconductivity, forced the total unit flux of the monopole to equally divide to half fluxons going along $\pm\hat{n}_0$. In this case (and ONLY for such a flux!) is the A.B. phase equal to $\pi$, independent of the $\pm\hat{n}_0$ orientation.

A simple topological argument for such a phase emerges when we ask what the A.B. phase is accumulated in a 2-d setup when a point vortex encircles a portion of the bound state of a quantum charge (electron). For the case of half fluxon, time reversal invariance allows only for a vanishing phase (e.g., for a path which does not encircle any charge) or a phase of $\pi$, e.g., for paths encircling all charges. Therefore there is some special point (and in general an odd number of such points) inside the wave function support, the encircling of which entails a jump of the phase by $\pi$. This is the point where a half fluxon stationed there make the ground state level in question[5] degenerate.

This indeed is the Berry phase for the case of two fold degeneracy. The case of just one degeneracy point at the origin of the 3-d space is special. The geometric phase is here the average phases of $\pi$ arising if the flux was directed along any one of the rays included inside the solid angle encircled, with total weight $\Omega$, and the 0 phase due to the rest, $4\pi - \Omega$ of the sphere.

Remarkably the E.M. analog tends to reconstruct the original spin-1/2 system. The magnetic and electric fields of a monopole at the origin and a charge at $\vec{R} = R\hat{n}$ generate angular momentum $\sim \int d^3\vec{r} \ \ \vec{r} \times (\vec{E} \times \vec{B}) \sim 1/2 \, eg\hat{n}$, where using Dirac's quantization[4](eg=1) we obtain a spin-1/2 pointing along the unit vector $\hat{n}$ between the monopole g and the charge e, precisely the $\hat{\sigma} \cdot \hat{n}$ we started from! (see fig. 4.)

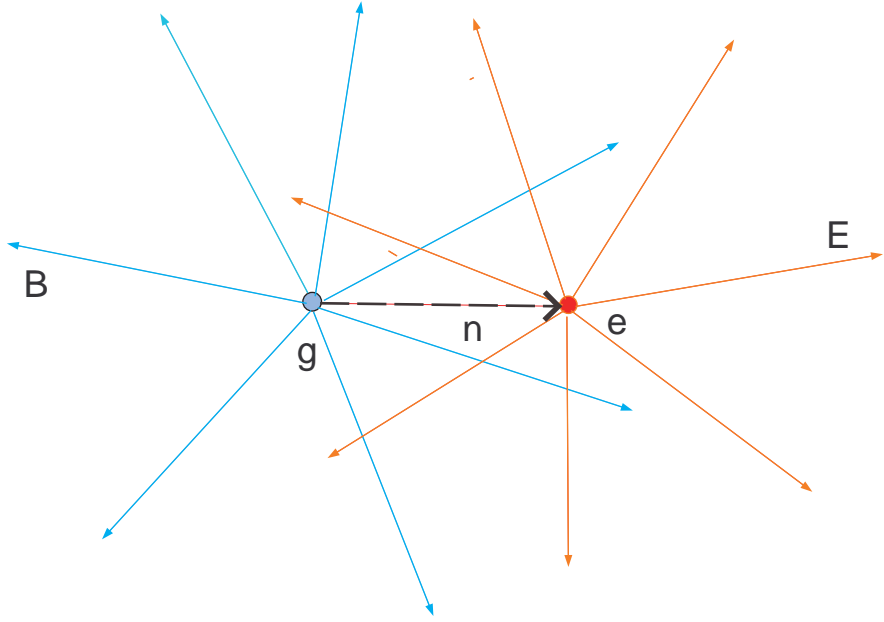

Fig. 4. E and B fields of magnetic monopole g and electric charge e

SU(2) → SU(N) extensions appear straightforward but are rather non-trivial. In the above SU(2) we had U(1) ∼ O(2) subgroup corresponding to rotations around the preferred $\hat{n}_0$ "Degeneracy" axis above (or around half fluxons along the $\pm\hat{n}_0$ direction in the E.M./A.B analog). Formally this subset of SU(2) was selected by time reversal symmetry and the need to avoid imaginary pieces in the Hamiltonian. The SU(3) analog of the Pauli matrices are the eight Gell-Mann $l_k$ in the lie algebra: $[l_k, l_m] = i f(kmn) l_n$; the $f(kmn)$ are real. Hence the three imaginary $l_k$ span a SU(2)= O(3) subgroup. The latter can be embedded in many ways in the SU(3)/O(3) manifold like in the choice of $\hat{n}_0$ above on the sphere.

Note that for imaginary $l_k$ the finite group elements $\exp i(\alpha_k \cdot l_k)$ are real. In general, SU(N) has upon restriction to real transformations O(N) sub-groups. The analog of the heavy probe at $R(t)$ in three-dimensional $R$ space is now a point in the $n^2 - 1$ dimensional parameter space $A_k(t) = 1...n^2 - 1$, with the $n$ level hamiltonian $H = A_k\, G_k$, with $G_k$ the $n^2 - 1$ generators in the fundamental representations the SU($n$). The infinite or close line degeneracy domain C* in the SU(2) case becomes here a closed (or infinite surface) S* of dimensionality $n(n+1)/2 - 1$ in the SU($n$)/O($n$) manifold. The topological Berry phases are now picked up when the system describes a closed path in $n^2 - 1$ dimensional parameter space which threads the degeneracy surface S*. These are "Genuine" $n$-fold degeneracies of *all n* levels corresponding to the complete vanishing of the Hamiltonian and cannot be reduced to lower degeneracies of, say, $n - 1$ out of the $n$ levels and trivial SU($n$) $\rightarrow$ SU($n$-1) reductions.

Next we introduce all $A_k$ components—in which case we can write just like before: $i\,\Sigma A_k/|\vec{A}| \cdot G_k$, with the sum extending over all the group generators. Interesting questions—not addressed here—relate to the analog of the geometric Berry phase and higher dimensional analog of the monopole's charges and angular momentum—now constructed presumably as exterior products of higher-order forms with many different coordinates:

How does the (ordinary 3-d) geometric Berry phase arise from the point of view of the atom/spin-1/2 system? We answer this for a particular probe trajectory, namely, an equatorial circle of radius $R$ traversed at constant (angular) velocity $(w)$ and speed $v = Rw$. This provides a simple illustration of the complementary point of view where small deviations from true adiabaticity integrate to the final, finite, geometric-like phases.

The Hamiltonian for this case is $H(t) = A[cos(wt)\sigma(1) + sin(wt)\sigma(2)]$. Transforming via $U = \exp(iwt\,\sigma(3)/2)$ to the "rotating frame", the Hamiltonian becomes $H' = U^+HU + iU^+\frac{\partial}{\partial t}U = A\,\sigma(1) + w\,\sigma(3)/2$. The new $H'$ implies a non-vanishing expectation value of $\sigma(3): <\sigma(3)> = w/(2A)$. This E.V. measures the (small) deviation from adiabaticity: $|A|$ is the level spacing and $w$ the rate of the probe's periodic motion. With $w/A = \epsilon << 1$ the system stays in its lowest state anti-aligned with the instantaneous $\hat{n}$ direction with a *crucial* tiny correction: It tracks it not in the equatorial plane (as it should in the perfectly adiabatic case) but is slightly shifted up to an almost flat cone of opening angle $\pi - \epsilon$! Since $\sigma(3)$ is invariant under the $U = \exp(i\,\sigma(3)wt)$ rotations the phase acquired due to it keeps accumulating. The (long!) time required for the probe to complete the cycle is $T = 2\pi/w$. The corresponding accumulated phase is then: $A\epsilon <\sigma> \cdot T = A(w/A)(2\pi/w) \cdot 1/2 = \pi$. Thus both the long period time and strong coupling $A$ cancel out and we are left with just the geometric phase corresponding to the solid angle of half a sphere! It is relatively simple to generalize this to smaller "latitude" circles at polar angle $\theta$ of radius $R\sin\theta$ enclosing a solid angle $\Omega = \pi \cdot (1 - \cos\theta) = 2\pi\,\sin^2\theta/2$. Finally, any part of the sphere can be exhausted by infinitely many such tangent circles generalizing the proof to any closed path.

# 6 Adiabatic complete MSW switching of solar neutrinos: a non-abelian Berry phase

To tie in SCINSP neutrinos consider adiabatic $\nu_e \to \nu_\mu$ MSW conversion in the sun. As a $\nu_e$ produced at the solar core moves outwards it encounters eventually a layer where matter effects make it degenerate with $\nu_\mu$. For an extended crossover region even tiny $\mu, e$ mixings cause a complete $\nu_e$ to $\nu_\mu$ conversion as in the famous figure in J. Bahcall's book.[6]

An entertaining classical analog I showed in a colloquium using a demo that Robert Sproul and George King arranged is the following: We start with two long weakly coupled "$e$" and "$\mu$" pendulums. Initially only the "$e$" pendulum is excited. It is a bit longer with lower natural frequency: $\Delta = (g/l_\mu)^{1/2} - (g/l_e)^{1/2} < 0$. A slow motor—mimicking the changing matter effects—adiabatically shortens it so that in the end the $\mu$ pendulum is longer and $\Delta$ flips sign. Since the (small!) coupling $\epsilon$ of the two pendulums stays constant, we have resonant beating with many back and forth energy transfers between the two pendulums.

Remarkably at the end when $|\Delta| >> \epsilon$ *all* the energy is transferred from one pendulum to the other, just as all the initial $\nu_e$ wave becomes a pure $\nu_\mu$ in the sun.

To explain this using the Berry phase we use the Hamiltonian: $H = \epsilon\sigma(1) + \Delta\sigma(3)$. The above experiments performed half a circuit in the $\Delta - \epsilon$ plane. Had we completed the circuit with $\Delta$ changing sign again regaining the initial configuration, then the system would return to the initial pure "$e$" (i.e., only the "$e$" pendulum excited) state possibly with a 1/-1 overall topological Berry phase[7]. The second lengthening stage is—by symmetry—the same as the first. Hence the transformation $T$ in the $e - \mu$ Hilbert space corresponding to half a circuit satisfies $T^2 = -1$ implying that $T$ effects a "$\mu$" $\leftrightarrow$ "$e$" transposition[7].

# Acknowledgement

The subject of my presentation differs from most others (Freedman's excepted!). Yet it closely ties with another, less familiar, aspect of Frank: his passion for and contributions to theoretical physics. This manifested not only in the theoretical papers that he has coauthored. Rather, Frank got Yakir Aharonov, the late Jeeva Anandan (whose true passion were these phases), Pawel Mazur, and the "Quantum Summer Institute". Thanks to him I have been visiting the University of South Carolina for the last 22 years—a joyful experience for which I am truly grateful.

[7] To close the path of an elongated rectangular shape in the $\Delta-\epsilon$ plane, we need to change $\epsilon \to -\epsilon$ for, say, $\Delta > 0$ and $\Delta \gg \epsilon$ and again -$\epsilon \to \epsilon$ at the opposite point $\Delta < 0$ but $|\Delta| \gg |\epsilon|$. Since this is done far away from resonance (namely, $\Delta = 0$), no further change in the pendulum's state ensues.